# Electronic transport computation in thermoelectric materials: From ab initio scattering rates to nanostructures

Neophytos Neophytou[1*], Pankaj Priyadarshi[1], Zhen Li[1], and Patrizio Graziosi[2]

[1] School of Engineering, University of Warwick, Coventry, CV4 7AL, UK
[2] Institute of Nanostructured Materials, CNR, Bologna, Italy

[*]N.Neophytou@warwick.ac.uk

# Abstract

Over the last two decades a plethora of new thermoelectric materials, their alloys, and their nanostructures were synthesized. The *ZT* figure of merit, which quantifies the thermoelectric efficiency of these materials increased from values of unity to values consistently beyond two across material families. At the same time, the ability to identify and optimize such materials, has stressed the need for advanced numerical tools for computing electronic transport in materials with arbitrary bandstructure complexity, multiple scattering mechanisms, and a large degree of nanostructuring. Many computational methods have been developed, the majority of which utilize the Boltzmann transport equation (BTE) formalism, spanning from fully ab initio to empirical treatment, with varying degree of computational expense and accuracy. In this paper we describe a suitable computational process that we have recently developed specifically for thermoelectric materials. The method consists of three independent software packages that we have developed and: 1) begins from ab initio calculation of the electron-phonon scattering rates, 2) to then be used within a Boltzmann transport simulator, and 3) calculated quantities from BTE are then passed on to a Monte Carlo simulator to examine electronic transport in highly nanostructured material configurations. The method we describe is computationally significantly advantageous compared to current fully ab initio and existing Monte Carlo methods, but with a similar degree of accuracy, thus making it truly enabling in understanding and assessing thermoelectric transport in complex band, nanostructured materials.

**Index terms:** thermoelectricity, complex bandstructure materials, ab initio scattering rates, Boltzmann transport, Monte Carlo, computational methods, nanostructures.

# I. Introduction

Thermoelectric generators (TEGs) are solid-state devices able to convert the heat flow arising from temperature gradients directly into electricity. They have the potential to offer a sustainable path for power harvesting from a variety of industrial sectors at power levels from microwatts to tens/hundreds kW, and even MW. Their impact could be widespread across many applications including medical, wearable electronics, building monitoring, the internet of things, refrigeration, thermal management, space missions, transportation, and various industrial sectors [1]. However, high prices, toxicity, scarcity, and low efficiencies of the prominent thermoelectric (TE) materials are currently hampering their large-scale exploitation. On the other hand, progress on these materials has been rapidly expanding over the last two decades. Novel concepts and improved understanding of materials synthesis have provided new opportunities for the enhancement of the thermoelectric conversion efficiency across many materials [1]. The thermoelectric figure of merit $ZT$, which quantifies the ability of a material to convert heat into electricity, has more than doubled compared to traditional values of $ZT$~1, reaching values of $ZT > 2$ in several instances across materials and temperature ranges, and even approaching 3 in some cases [1, 2, 3, 4, 5].

The TE performance is quantified by the $ZT$ figure of merit as $ZT = \sigma S^2 T/(\kappa_e + \kappa_l)$, where $\sigma$ is the electrical conductivity, $S$ is the Seebeck coefficient, $T$ is the absolute temperature, and $\kappa_e$ and $\kappa_l$ are the electronic and lattice parts of the thermal conductivity, respectively. The product $\sigma S^2$ in the numerator of $ZT$ is called the power factor (PF). The recent improvements in $ZT$ are mostly attributed to drastic reductions of the lattice thermal conductivity in nanostructured materials and nanocomposites, which has reached amorphous limit values at $\kappa_l = $ 1-2 W/mK and below [1, 6, 7, 8, 9]. One of the most successful strategies to reduce thermal conductivity is to hierarchically nanostructure the materials, by introducing different types and sizes of nano-features, which would scatter more effectively different groups of phonon mean-free-paths.

Nanostructuring has brought tremendous success, but it is gradually running out of steam in providing further reductions to thermal conductivity. It is becoming increasingly clear that any further benefits to $ZT$ must now come from power factor improvements.

Current research efforts in improving the power factor aim towards identifying materials with favorable bandstructure features, such as resonant states and low-dimensional 'like' features within bulk materials [10, 11], or bandstructure engineering such as band-convergence strategies [11, 12, 13, 14]. Indeed, most promising TE materials have complex bandstructures, with multiple valleys and bands that extend in the entire Brillouin Zone (BZ).

Due to the vast material possibilities, the exploration for PF enhancing band features needs to be led by advanced computational studies, which elucidate their role in electronic and TE transport. The most common computational approach is to extract the electronic bandstructures using ab initio methods, and afterwards use the Boltzmann Transport Equation (BTE) to extract the TE coefficients. For the scattering rates used within the BTE the most common approach is the constant relaxation time (CRT) or constant mean-free-path (CMFP) approximations, such as using *BoltzTrap* code [15], where a CRT around 10 fs, or a CMFP of around 5 nm, are routinely employed [15, 16]. These approximations have the advantage of being computationally efficient, but have the disadvantage of providing uncertain and rather arbitrary outcomes, both quantitatively and qualitatively, e.g. with respect to materials ranking, temperature trends, and optimal carrier density [17, 18]. However, we know that in these materials, the electronic scattering processes have complex energy, momentum and band dependencies [18].

More elaborate computational methods for the treatment of electron-phonon (e-ph) scattering vary, depending on the required accuracy and computational complexities, i.e. by only considering the energy dependence of acoustic phonons analytically [19]; by considered elastic scattering by acoustic phonons in a full band approach using deformation potentials [20]; by extracting deformation potentials from band shifts after applying stress along specific directions and using those to form scattering rates [21]; by using a full band approach and a numerical scheme for electron-phonon scattering based on calculating the ab initio electron-phonon coupling matrix elements, but with constant optical phonon frequencies [22, 23]; and by full ab initio EPW+Wannier e-ph scattering rate calculations using the EPW code [24]. Other than for 2D materials, it is computationally challenging to utilize the entire phonon and electron dispersions for full transport calculations, thus, it is common to employ reasonable approximations [23]. Overall, however, the majority of current TE simulation works (justifiably) tend to heavily sacrifice accuracy for computational speed.

The need for theory to lead experiment, the search for new materials, and detailed understanding of the transport physics that would lead to performance optimization, requires electronic transport tools that provide confidence in accuracy, as well as computational robustness. We have recently developed such tools, which fill the gap of higher accuracy, in the inevitable expense of moderate computational cost, but still much lower compared to fully ab initio simulations. Our method is based on ab initio calculations of a selected few electron-phonon matrix elements, out of which we form deformation potentials and scattering rates based on the deformation potential theory. We then use those within our newly developed full-band BTE code, *ElecTra*, which provides the electronic properties of the bare material. This information is then passed on to a third simulator, a Monte Carlo real space ray-tracing code, specifically designed to address challenges for simulating TE transport in nanostructures. These three main sections are presented separately in the detailed works in Refs. [25, 26, 27], but in this paper we provide an overview of the entire process.

The paper is organized as follows. In Section II we describe the BTE code. In Section III we describe the extraction of the deformation potentials from ab initio calculations. In Section IV we present how we use the BTE and ab initio information in the MC code and present an example for transport simulations in a nanostructured material. Finally, in Section V we conclude.

## II. Boltzmann transport using the *ElecTra* code

To evaluate the transport properties of the base material, we use our newly developed simulator, *ElecTra*, that considers all appropriate scattering mechanisms (acoustic phonon, optical phonon, ionized impurity scattering), and the full energy, momentum, and band dependence of the relaxation times. Here we provide a few details on the process flow of the code, whereas more details can be found in the dedicated *ElecTra* paper and the user's manual [26, 28].

**Linearized Boltzmann Transport Equation (BTE) formalism**

The TE transport coefficients are extracted within the Linearized Boltzmann Transport Equation (LBTE) formalism under the relaxation time approximation as [29, 30]:

$$\sigma_{ij(E_F,T)} = q_0^2 \int_E \Xi_{ij}(E) \left(-\frac{\partial f_0}{\partial E}\right) dE, \quad (1a)$$

$$S_{ij(E_F,T)} = \frac{q_0 k_B}{\sigma_{ij}} \int_E \Xi_{ij}(E) \left(-\frac{\partial f_0}{\partial E}\right) \frac{E-E_F}{k_B T} dE, \quad (1b)$$

$$\kappa_{e\,ij} = \frac{1}{T} \int_E \Xi_{ij}(E) \left(-\frac{\partial f_0}{\partial E}\right) (E-E_F)^2 dE - \sigma S^2 T \quad (1c)$$

where $\Xi_{ij}(E)$ is the Transport Distribution Function (TDF) defined below in Eq. (2), $E_F$, $T$, $q_0$, $k_B$, and $f_0$, are the Fermi level, absolute temperature, electronic charge, Boltzmann constant, and equilibrium Fermi distribution, respectively.

The TDF is expressed as a surface integral over the constant energy surfaces, $\mathfrak{L}_E^n$, for each band, and then summed over the bands, as [29, 30, 31]:

$$\Xi_{ij(E,E_F,T)} = \frac{s}{(2\pi)^3} \sum_{k,n}^{\mathfrak{L}_E^n} v_{i(k,n)} v_{j(k,n)} \tau_{i(k,n,E_F,T)} \frac{dA_{k_{\mathfrak{L}_E^n}}}{|\vec{v}_{(k,n)}|} \quad (2)$$

where $k_{\mathfrak{L}_E^n}$ is a state on the surface $\mathfrak{L}_E^n$ and $dA_{k_{\mathfrak{L}_E^n}}$ is its corresponding surface area element, computed as explained below below. $v_{i(k,n)}$ is the *i*-component of the band velocity of the transport state, $\tau_{i(k,n)}$ is its momentum relaxation time, $\frac{dA_{k_{\mathfrak{L}_E^n}}}{|\vec{v}_{(k,n)}|}$ is its density-of-states (DOS), and *s* is the spin degeneracy.

The relaxation times for each individual scattering mechanism are combined following using Matthiessen's rule for each $(k,\mathfrak{L}_E^n)$ state, to compute the comprehensive TE coefficients. Also, the overall energy-dependent relaxation time $\tau$ and mean-free-path (mfp) $\lambda$ are returned, both per-band, per scattering mechanism, and overall for all mechanisms. These are computed as:

$$\lambda_{i(E,E_F,T)} = \frac{\sum_{k,n}^{\mathcal{E}_E^n} |v_{i(k,n,E)} \tau_{i(k,n,E,E_F,T)}| DOS_{(k,n,E)}}{\sum_{k,n}^{\mathcal{E}_E^n} DOS_{(k,n,E)}} \tag{3a}$$

$$\tau_{i(E,E_F,T)} = \frac{\sum_{k,n}^{\mathcal{E}_E^n} \tau_{i(k,n,E,E_F,T)} DOS_{(k,n,E)}}{\sum_{k,n}^{\mathcal{E}_E^n} DOS_{(k,n,E)}} \tag{3b}.$$

These are meaningful full-band transport quantities, to be further exploited in subsequent simulations (such as Monte-Carlo).

**Electronic structure quantities**

The first step in the simulation approach is to obtain the electronic structure using Density Functional Theory (DFT). *ElecTra's* interface can take as input a '.bxsf' format file [32, 33] enabling to interface with any DFT code that provides the electronic structure data in this format. This is the format used by 'XCrysDen' [33], which is compatible with a number of DFT codes. For example, QE, CASTEP and VASP among others, have the fs, c2x, and vasp2x_fs routines, respectively, to save the computed DFT bandstructure in this format.

Then *ElecTra* builds the constant energy surfaces to grasp the 3D details of the bandstructure. For this, we map the function $E(\mathbf{k})$ into a function $\mathbf{k}(E)$. The 3D mesh in the $k$-space is scanned, the mesh elements crossed by the constant energy surfaces are identified, and the coordinates of the points on these surfaces are computed. For every element of the $\mathbf{k}(E)$ mesh, the $E(\mathbf{k})$ is linearly interpolated from the original mesh, and the three components of the band velocity $v_{i(\mathbf{k},n,E)}$ are computed with the contragradient method [34]. *ElecTra* offers two schemes to perform this step: i) A local triangulation on the $k$-space mesh elements which are crossed by the surface of the constant energy of interest. Here the elemental surface area, $dA_{(\mathbf{k},n,E)}$, which will define the DOS of that specific state, is computed using Heron's formula. ii) An easier approximate method which samples the nearest-neighbour points on the $k$-mesh (NN-scan). Here the code scans all the $\mathbf{k}$-points on the energy surface, checks for their nearest neighbours, then uses the distances between these $\mathbf{k}$-points to approximate the d$A_k$ surface element areas. In the latter, *ElecTra* does not resolve constant energy surface elements, but acquires a collection of points on the energy surface of interest. Although this is an approximation, as it detects only the points along the edges of the *k*-mesh elements, it is around 15 to 30 times faster, but without noticeable penalty in the results compared to triangulation, either for isotropic or anisotropic bands. For further details about these methods we refer the interested reader to the original Electra paper and the user's manual [26, 28].

The *k*-state-dependent DOS is then defined as $\frac{dA_{(k,n,E)}}{|\vec{v}_{(k,n,E)}|}$, and used later in the scattering rate calculations as well as the energy integrations. For each band in the electronic structure, the energy-dependent DOS is then calculated as:

$$\text{DOS}(E,n) = \oiint_{\mathfrak{L}_E^n} \frac{dA_{(k,n,E)}}{|\vec{v}_{(k,n,E)}|} = \frac{s}{(2\pi)^3} \sum_{k_{E,n}} \frac{dA_{(k,n,E)}}{|\vec{v}_{(k,n,E)}|} \qquad (4)$$

where *s* is the spin degeneracy taken as 1 or 2, and $\vec{v}_{(k,n,E)}$ is the band velocity. The comprehensive DOS(*E*) is the sum of the DOS of all individual bands and the comprehensive velocity $v(E)$ is the average of the state velocity $v(E,n) = \langle |\vec{v}_{(k,n,E)}| \rangle_k$. Note that we have used energy surface elements that are extracted after we construct the constant energy surfaces, which is an alternative way compared to codes such as *BoltzTrap* which uses volume integral and delta-functions represented by smeared Gaussians [15].

The method has been systematically validated with details reported in the manual [28] for various example cases: i) comparisons with parabolic and non-parabolic bands for which the DOS, velocity and TDF solutions for isotropic scattering mechanisms are analytical, ii) comparisons for the mobility of materials cases for which the mobility is experimentally well-known, i.e. Si, Ge, SiGe, and GaAs, and iii) comparisons with quantities (i.e. DOS) extracted from existing codes. The latter is depicted in Fig. 1 for TiCoSb and $Mg_3Sb_2$.

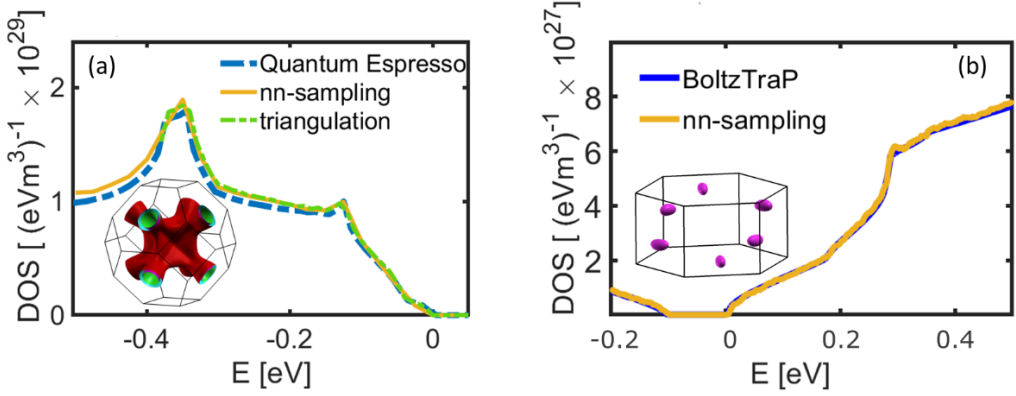

**Figure 1:** (a) Comparison of the DOS of the valence band of TiCoSb using Quantum Espresso and *ElecTra*, the latter from the two methods implemented (nn-sampling and triangulation). The inset shows constant energy surfaces of one of the valence bands at $E = -0.12$ eV below the valence band edge. (b) DOS for the conduction band of $Mg_3Sb_2$, and comparison between the results obtained from *BoltzTraP* [15] and *ElecTra* [26]. The inset depicts the constant energy surface for the $Mg_3Sb_2$ conduction band at $E = 0.1$ eV above the band edge.

**Scattering rates and transport:**

For each transport state ($k$,$n$,$E$) and each scattering mechanism $m_s$, the corresponding momentum relaxation time $\tau_{i(k,n,E)}^{(m_s)}$ is defined from Fermi's golden rule as:

$$\frac{1}{\tau_{i(k,n,E)}^{(m_s)}} = \frac{1}{(2\pi)^3} \sum_{k'} |S_{k,k'}^{(m_s)}| \left(1 - \frac{v_{i(k')}}{v_{i(k,n,E)}}\right) \quad (5)$$

where the sum runs over all the allowed final states $k'$ of the same carrier spin [30, 31]. $|S_{k,k'}|$ is the transition rate between the initial $k$ and final $k'$ states, computed as detailed by Eq. 6 below for the different mechanisms. The $\left(1 - \frac{v_{i(k')}}{v_{i(k,n,E)}}\right)$ term is an approximation for the momentum relaxation time, which is used to solve the BTE in the closed form, as commonly done in the literature when computing the transport coefficients [35]. *ElecTra* computes the scattering rates for the different scattering mechanisms as [30, 31]:

$$\left|S_{k,k'}^{(ADP)}\right| = 2\frac{\pi}{\hbar} D_{ADP}^2 \frac{k_B T}{\rho v_S^2} g_{k'} \quad (6a)$$

$$\left|S_{k,k'}^{(ODP)}\right| = \frac{\pi D_{ODP}^2}{\rho\omega} \left(N_\omega + \frac{1}{2} \mp \frac{1}{2}\right) g_{k'} \quad (6b)$$

$$\left|S_{k,k'}^{(IVS)}\right| = \frac{\pi D_{IVS}^2}{\rho\omega} \left(N_\omega + \frac{1}{2} \mp \frac{1}{2}\right) g_{k'} \quad (6c)$$

$$\left|S_{k,k'}^{(POP)}\right| = \frac{\pi q_0^2 \omega}{|\mathbf{k}-\mathbf{k'}|^2 \varepsilon_0} \left(\frac{1}{k_\infty} - \frac{1}{k_s}\right) \left(N_\omega + \frac{1}{2} \mp \frac{1}{2}\right) g_{\mathbf{k'}} \tag{6d}$$

$$\left|S_{k,k'}^{(IIS)}\right| = \frac{2\pi}{\hbar} \frac{Z^2 q_0^4}{k_s^2 \varepsilon_0^2} \frac{N_{imp}}{\left(|\mathbf{k}-\mathbf{k'}|^2 + \frac{1}{L_D^2}\right)^2} g_{\mathbf{k'}} \tag{6e}$$

$$\left|S_{k,k'}^{(Alloy)}\right| = \frac{2\pi}{\hbar} \Omega_c x(1-x) G_{\mathbf{k}-\mathbf{k'}} \Delta E_G^2 g_{\mathbf{k'}} \tag{6f}$$

Above, ADP refers to 'Acoustic Deformation Potential' and is the scattering between charge carriers and acoustic phonons. ODP stands for 'Optical Deformation Potential' and describes the charge carrier inelastic scattering with non-polar optical phonons. Both can be chosen to be intra- and/or inter-*band*. IVS stands for 'Inter-Valley Scattering' and it is specific for the inelastic inter-valley scattering. POP stands for 'Polar Optical Phonon' and describes the inelastic/anisotropic scattering of charge carriers with polar phonons, which here is treated as both intra- and inter-band [30]. *ElecTra* allows different phonon frequencies for all these inelastic processes separately, for example, for each non-polar and polar phonon branches, different frequencies can be used. IIS stands for 'Ionized Impurity Scattering' and describes the elastic scattering rate due to ionized dopants, for which the user can choose both intra- and/or inter-band transitions. "Alloy" represents the alloy scattering due to intrinsic disorder in alloys or solid solutions and is considered both intra- and inter-band [35]. ***k*** and ***k'*** are the wave vectors of the initial and final states. "-" and "+" in Eqs. 6b-7d indicate the phonon absorption and emission processes, respectively. These type of transition processes are illustrated in Fig. 2a. We allow for the directional dependence of the momentum scattering rates, because in an anisotropic band and /or under the influence of anisotropic scattering mechanisms, the rate at which the carrier's momentum relaxes, will depend on the carrier's initial momentum and the distribution of momenta of the final states.

The variables that appear in Eq. 6 are as follows: $D_{ADP}$, $D_{ODP}$, $D_{IVS}$ are the deformation potentials for the ADP, ODP, and IVS mechanisms. $\rho$ is the mass density, $v_s$ the sound velocity, $\omega$ the dominant frequency of optical phonons, considered as constant over the whole reciprocal unit cell, which has been validated to be a satisfactory approximation, [22] and $N_\omega$ is the phonon Bose-Einstein statistical distribution; $\varepsilon_0$ the vacuum dielectric constant, $k_s$ and $k_\infty$ the static and high frequency relative permittivities, $Z$ the electric charge of the ionized impurity considered, and $N_{imp}$ is the density of the ionized impurities. $g_{k'} = \frac{dA_{(k',n,E)}}{|\vec{v}_{(k',n,E)}|}$ is the single-spin DOS of the final scattering state. $L_D = \sqrt{\frac{k_s \varepsilon_0}{e} \left(\frac{\partial n}{\partial E_F}\right)^{-1}}$ is the generalized screening length with $E_F$ being the Fermi level and $n$ the carrier density [29, 30]. $\Omega_c$ is the volume of the primitive cell, $x$ the fraction of one of the alloy elements, and $\Delta E_G$ the difference between the energy gap of the two constituent materials that form the alloy. The $G$ function is the form factor of a hard sphere [35].

The scattering rates and the transport coefficients are computed along the orthogonal Cartesian space directions $x$, $y$, $z$. Consequently, the constant energy surfaces are expressed in Cartesian coordinates on orthogonal axes instead of unit cell axes, and the reciprocal unit cell is used instead of the Brillouin Zone, as depicted in Fig. 2b.

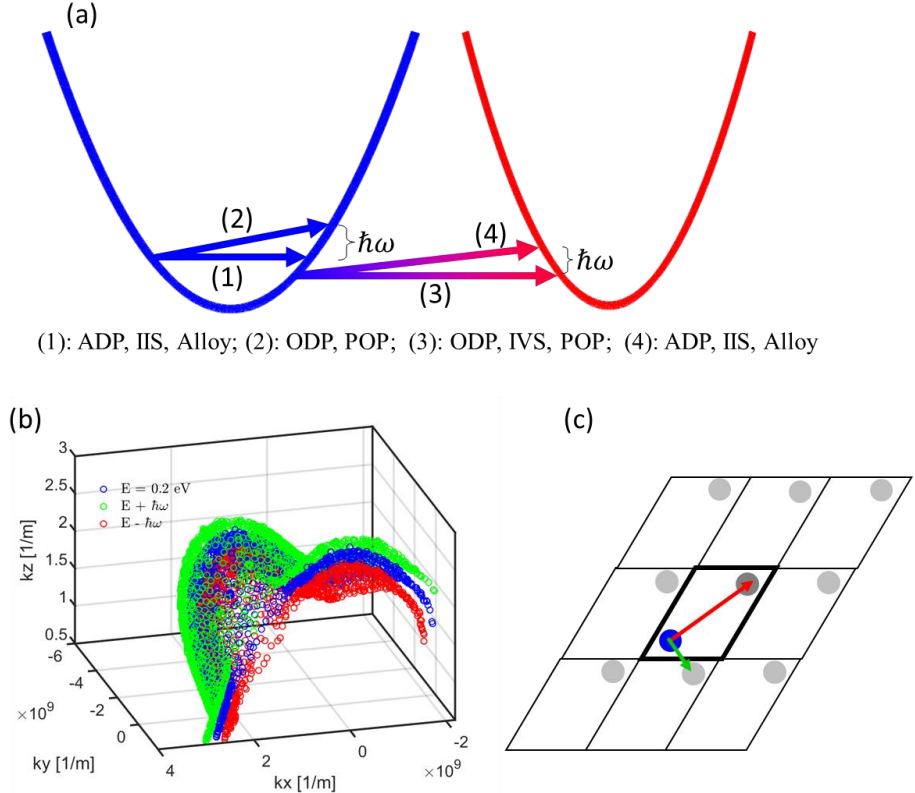

**Figure 2**: (a) Schematic of two bands with the four types of allowed transitions: (1) elastic intra-band; (2) inelastic intra-band, (3) inelastic inter-band, (4) elastic inter-band. $\hbar\omega$ is the energy of the absorbed or emitted phonon. (b) Slice view of three constant energy surfaces from a valence band of ZrNiPb in the reciprocal unit cell, each separated by the phonon energy (with the blue coloured surface in the middle). Each depicted point is an actual transport state detected and used by *ElecTra*. An inelastic scattering involving absorption/emission transitions from the blue surface will result in a final state on the green/red surfaces. (c) Illustration of a 2D schematic of reciprocal unit cell with bold edges, and its equivalent cells around it. The initial *k*-point is in blue and the final *k*-point in dark grey. The equivalent final *k*-points in an extended zone scheme are shown in fainted grey, obtained by translating the blue point by one reciprocal lattice vector in all possible directions. In the anisotropic POP and IIS scattering mechanisms, all the equivalent *k*-points are explored, and the actual final *k*-point considered is the one closest to the initial *k*-point. Thus, the exchange vector considered by *ElecTra* in the depicted example is the green one and not the red one.

A further important point is that the POP and IIS scattering strengths depend on the momentum scattering vector, i.e. the distance in the *k*-space between the initial and final states $|\boldsymbol{k} - \boldsymbol{k}'|^2$. To compute this exchange vector, *ElecTra* considers every final state's $\boldsymbol{k}$-point position in all neighbouring reciprocal unit cells, and then uses the minimum distance from the initial point in the exchange vector. A 2D schematic of this process is depicted in Fig. 2c. The final point in the reciprocal unit cell (dark grey bullet), is shifted by the reciprocal lattice vectors in all possible directions, creating the fainted grey bullet points. The central cell in Fig. 2c with bold edges represents the cell used in the simulations and the other cells are the equivalent ones. Then the code considers the closest distant to the initial blue bullet point. This is necessary, because for example two $\boldsymbol{k}$-points that are located at opposite edges of the reciprocal unit cell are actually very near if the equivalent point in the nearest neighbour cell. Thus, in Fig. 2c the physical scattering vector is the green one, and not the red one.

*__Illustrative examples for realistic materials – the case of ZrNiPb:__* We now show the cases of relaxation times and mean-free-paths for the half-Heusler ZrNiPb in Fig. 3, for three different temperatures, with the scattering parameters used taken from Ref. [36]. We start with ADP and ODP. In Fig. 3a-b we show the relaxation time $\tau$ and in Fig. 3c-d the mean free path $\lambda$. For both cases we have the rise of $\tau$ and $\lambda$ at the band edge and reduction further into the bands. Note that these quantities are created based on deformation potential theory, but as of this point we have not performed a full validation comparison to full EPW+Wannier interpolated e-ph calculations for this material. We have performed such comparison for Si, for both n-type and p-type, using a full set of deformation potentials that we have extracted, with very good agreement – see Fig. 5 below [25].

In Fig. 3e-f, we show the relaxation time and mean-free-path when we consider IIS as well as ADP and ODP, for a donor doping density $n =$ of $1.1 \times 10^{20}$ cm$^{-3}$. Note that in this case we consider the effect of this IIS donor scattering on both the majority electrons and minority holes [29].

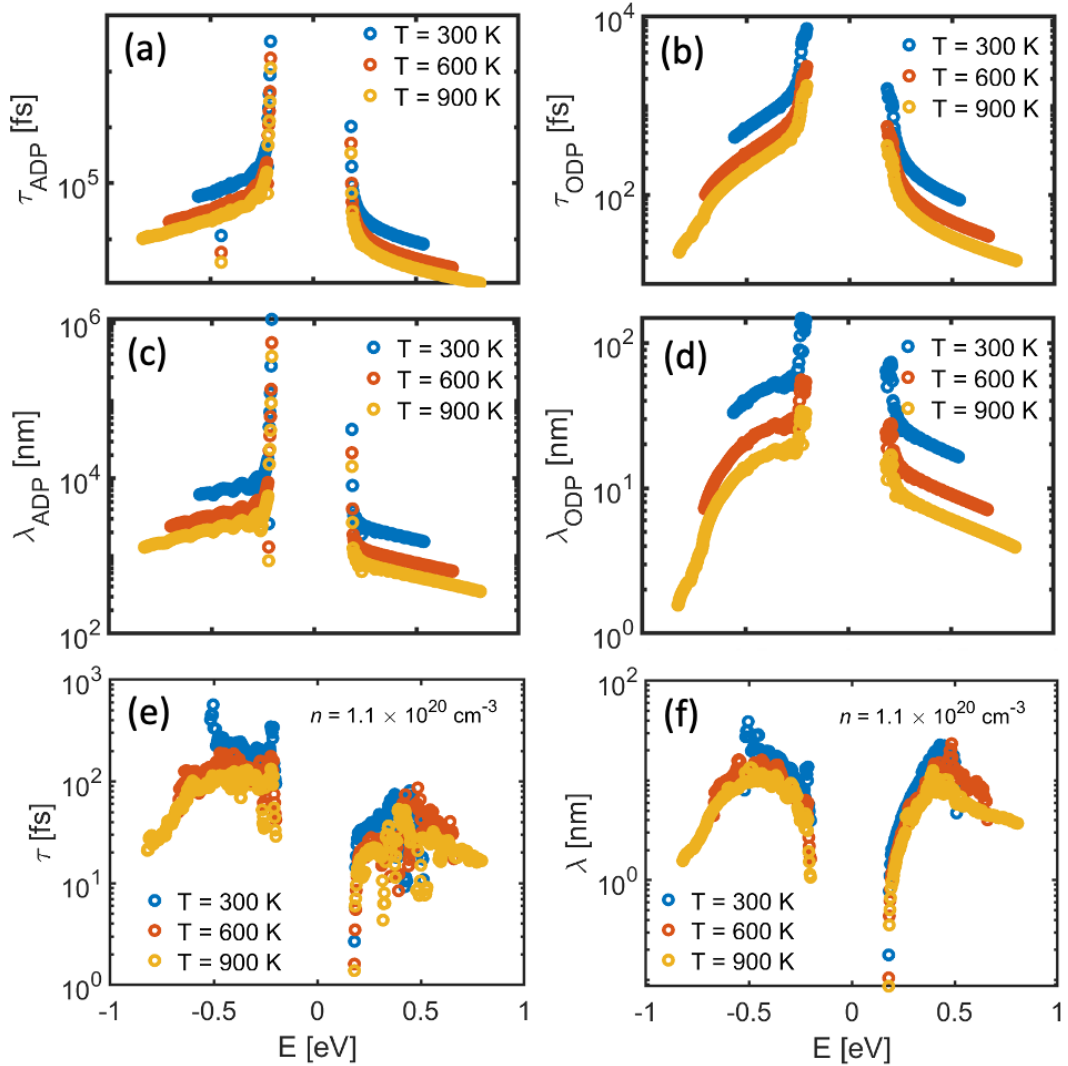

**Figure 3**: (a-b) Scattering times and (c-d) mean-free-paths for ZrNiPb. Quantities for ADP are in (a, c). Quantities for ODP are in (b, d). (e) and (f) show the relaxation times and mean free paths, respectively when IIS is considered on top of ADP and ODP.

***TDF illustrative examples:*** We now show examples of transport distribution functions (TDF) $\Xi(E)$, which contain all the information relevant to charge transport within the BTE.

We first show examples of these functions for different scattering mechanisms in the case of parabolic bands in Fig. 4a. ADP results in a linear $\Xi(E)$ function in this case, while for ODP the onset of the phonon emission at the characteristic phonon energy is clearly observed (here we used 50 meV). The comparison with the results from *ElecTra* validate the approach. In Figs. 4b-c we show the corresponding TDFs for ZrNiPb for ADP and ODP and three different temperatures, to indicate how a full-band treatment of complex bandstructures impacts the transport distribution functions. $\Xi(E)$ is no longer linear for ADP, and the abrupt jump at the phonon emission onset in the ODP case (Fig. 4c) is mitigated, while the fine features are smeared out as well. Note that the assumed optical phonon frequency is 24 meV. In Fig. 4d we consider the IIS on top of ADP and ODP using a donor impurity density of $1.1 \times 10^{20}$ cm$^{-3}$. The $\Xi(E)$ decrease at higher energies, especially after 0.5 eV inside the valence band, is driven by a DOS increase (black-solid line in Fig. 4d), which leads to an increase in scattering rate and a decrease in band velocity (black dashed line).

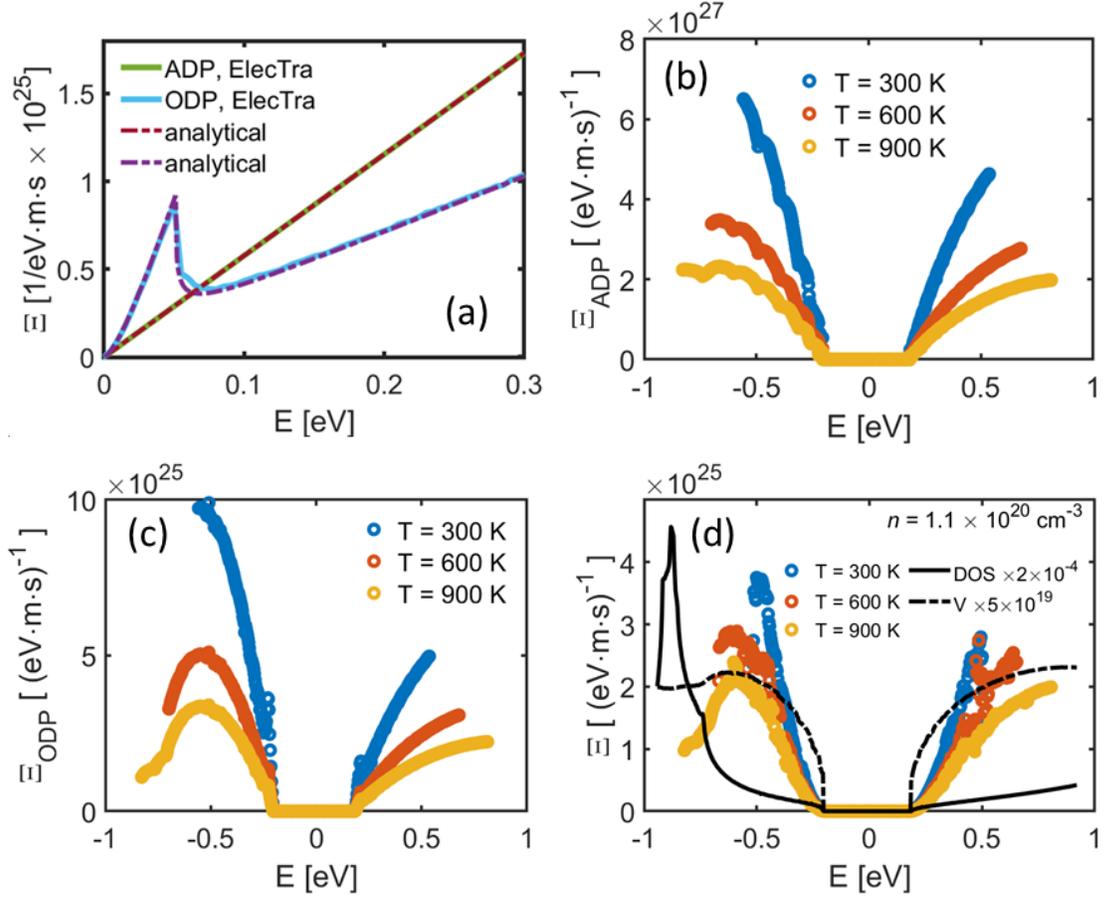

**Figure 4**: (a) The *xx* component of the TDF for an isotropic parabolic band ($m^* = 0.1$) for two different scattering mechanisms, comparing the analytical solution, dashed lines, and the ones computed numerically with *ElecTra*. Solid lines. (b), (c) and (d) show the *xx* component of the TDF for ZrNiPb, for ADP, ODP and ADP plus ODP plus IIS altogether, respectively. In (d), the DOS and band velocity *v* are shown in black solid and dash-dot lines, respectively, where an ad-hoc scaling has been performed for illustration purpose.

# III. Ab initio extraction of deformation potentials for scattering

The main scattering parameters required by *ElecTra* as described above, are the deformation potentials for electron-phonon scattering. Electron-phonon (e-ph) scattering is a vital part of simulations for materials properties. Deformation potentials are based on a theory developed by Bardeen and Shockley [37], and provide a convenient way to treat e-ph scattering using the analytical expressions in the previous section [30, 38]. The deformation potential essentially describes the shift in the bands upon a change in the lattice caused by a perturbation from specific phonon modes, the ones that dominate the overall process.

Deformation potential theory is instrumental for the calculation of low-field mobility [35, 39]. Traditionally, e-ph scattering is employed within transport methods such as the Boltzmann transport equation (BTE) [40, 41], Monte Carlo [42, 43], Landauer-Buttiker [44], etc.. It is also routinely used in semiconductor device transport simulators and high-field calculations in such devices, still with adequate accuracy [45, 46, 47]. The analytical scattering rates that result, are convenient when the e-ph scattering needs to be combined with other scattering mechanisms, such as for nanostructured materials [48], or highly doped materials and alloys for which ionized impurity scattering [18] and alloy scattering are important (such as transistor devices and thermoelectric materials [29]). The use of deformation potentials can allow for the flexibility and computational robustness that this process requires.

However, deformation potential values are only known for common semiconductor materials, and those are mostly extracted from experiment. For the arbitrary complex bandstructure materials, these are not known and need to be derived from first principles. In general, ab initio calculations are becoming a critical component in enabling the investigation of e-ph scattering processes. Although these works started in the 1980s [49, 50, 51], they are only recently expanded to complex materials as a result of the advancements in computational resources and software development. Recent advancements in ab initio methods and density-functional perturbation theory (DFPT) [52, 53, 54] theoretical methods and available software can now compute the electronic interactions from the entire phonon spectrum for materials beyond common semiconductors and solve the BTE using accurate scattering rates extracted from the calculation of all matrix elements [22, 24, 55]. They enable more accurate calculations without the use of empirical parameters and are used to predict and analyze new materials' transport properties, explore reaction mechanisms, and provide understanding in experimental synthesis and characterization.

Such approaches, however, are computationally extremely expensive, starting from their DFPT part. This can be typically performed on coarse meshes, but it is then interpolated using Wannier methods, because a dense electronic and phononic mesh discretization is required [56]. This leads to a large number of possible combinations of electronic and phononic states in the calculation of the e-ph interaction. For example, the use of interpolated 50×50×50 k-mesh and 25×25×25 q-mesh (typically half the size of the k-mesh), results in billions of possible matrix elements (the total number is given by the product of the two meshes). In fact, even larger meshes can be needed for mobility convergent calculations as shown in Ref. [58]. Typically, the symmetry of the unit cell is considered, which would reduce this number down to several million depending on the material specifics. Although these matrix elements are not calculated, but interpolated from the original coarse mesh onto the fine e-ph Wannier mesh [57], where they are subsequently used in transport calculations, this step is still computationally expensive. It can be even more expensive than the original DFPT/DFT on the coarse mesh (see Table 1 and discussions later on for a comparison example).

Here we describe a first-principles framework to extract deformation potentials in complex band materials based on density-functional theory (DFT) and density-functional perturbation theory (DFPT). We show that with the calculation of a reduced set of matrix elements and the formation of deformation potential scattering rate expressions rather than by using the full ab initio e-ph calculation, we can reduce the computational cost significantly, while not jeopardizing the accuracy [25]. We first compute the electronic band structures, phonon dispersion relations, and electron-phonon matrix elements to extract deformation potentials for acoustic and optical phonons for all possible processes. The matrix elements clearly show the separation between intra- and inter-valley scattering and quantify the strength of the scattering events. The deformation potentials are extracted and then be used within the BTE as described in the previous section.

The process is as follows: Using *ab initio* self-consistent field DFT calculations we extract the electronic structures of the material of interest [58, 59]. We employ density-functional perturbation theory (DFPT) [49, 50] calculations, which are based on the perturbative expansion of the Kohn-Sham energy functional, allowing calculations of vibrational properties within the DFT framework. Many electronic structure simulation packages implement and use DFPT, such as Quantum Espresso [53], ABINIT [54], VASP [60], etc. The use of DFT and DFPT calculations provides electronic structures, phonon dispersions, electron-phonon (*e-ph*) matrix elements, as well as $k_\infty$ and $k_S$, entirely from first principles. The key item is the determination of the *el-ph* coupling matrix elements, $M_{mn}^\nu$ which measure the coupling strength of the *el-ph* interaction, resulting in a transition process where an electron with wavefunction $\psi_{n\mathbf{k}}(r)$ at a state with band index *n* and wavevector **k** scatters into a state $\psi_{m\mathbf{k+q}}(r)$ with band index *m* and wavevector **k**+**q**. It is facilitated by an atomic displacement perturbation due to a phonon with mode index $\nu$ and crystal momentum **q**, and given by $M_{mn}^\nu(\mathbf{k}, \mathbf{q}) = \langle \psi_{m\mathbf{k+q}}(r) | \delta_{\nu q} V(r) | \psi_{n\mathbf{k}}(r) \rangle$, where $\delta_{\nu q} V(r)$ is the derivative of the self-consistent potential from DFPT. Codes which consider the variation formulation of DFPT provide the matrix elements, $g_{mn}^\nu(\mathbf{k}, \mathbf{q})$, which are the output of DFPT as $g_{mn}^\nu(\mathbf{k}, \mathbf{q}) = \sqrt{\hbar/(2m_0 \omega_{\nu q})} M_{mn}^\nu(\mathbf{k}, \mathbf{q})$ [61, 62], from which $M_{mn}^\nu$ can be extracted. Here $m_0$ is the sum of the masses of all atoms in the unit cell and $\omega_{\nu q}$ is the phonon frequency. Based on $M_{mn}^\nu$ for an individual transition, the deformation potential for acoustic and optical phonons can be derived as $D_\text{ADP} = M_{mn}^\nu(\mathbf{k}, \mathbf{q})/|q|$ and $D_\text{ODP} = M_{mn}^\nu(\mathbf{k}, \mathbf{q})$. $D_\text{ADP}$ is the slope of $M_{mn}^\nu$ and will require a few elements to be formed, whereas $D_\text{ODP}$ is directly the coupling matrix element. Deformation potentials in different directions can be extracted and then merged for simplicity. A larger number of matrix elements can be provided by Wannier interpolation (i.e. EPW code [24]) in a computationally efficient way, based on the initial direct DFPT calculations.

The $D_\text{ADP}$ (for LA and TA modes) computed in this way is comparable to the deformation potentials derived from band shifts after applying stress along specific directions, which is a good indication for an effective deformation potential, $D_\text{eff}$, as described recently in Ref. [21]. That method extracts the corresponding scattering rates based on $D_{\text{eff}}$, which is then used as a 'holistic' electron-phonon scattering process for all states and valleys in the bandstructure. This is indeed a very efficient way to extract first order transport properties at a very low computational cost, making it very effective form materials screening studies. However, it does not capture the details of the underlying physical mechanisms correctly, which can render it incomplete for TEs in certain cases. For example, it does not capture optical phonon scattering (important for nanostructuring), or inter-valley processes, crucial for band-alignment optimization, which is the main strategy for power factor improvements in complex materials. Our method provides a step forward in terms of accuracy and physical relevance. We distinguish between all acoustic and optical phonon scattering processes, and all intra- and inter-valley processes, accounting for all selection rules automatically as well. The price we pay is the need for the initial coarse mesh expensive DFPT calculations, but overall our method is less expensive compared to full EPW+Wannier due to the reduced number of matrix elements required in the calculation of the transport properties.

After identifying the dominant phonon modes, a number of targeted calculations are performed to extract the deformation potentials for a given material. Figures 5a-c show an example of this process for Si. Only a small electronic energy region near the valence band minimum (VBM) and the conduction band maximum (CBM) contribute to transport (red regions in Fig. 5a). For intra-valley scattering, energy/momentum conservation dictates that only a small part of the phonon spectrum takes part as well (red regions in Fig. 5b). Figure 5c shows the coupling matrix elements for the LA (blue line) and LO (red line) phonons around the VBM. For the former, the slope provides a $D_{\text{ADP}}$ value and for the latter the value near Γ is a $D_{\text{ODP}}$ value, which after necessary post-processing (for example by averaging around high symmetry lines, singular value decomposition for degeneracies [25, 68], etc.), provided $D$ values and mobility that agree very well with measured data, as presented in Ref. [25]. This method inevitably provides information about the strength of each process (optical/acoustic branches), as well as a clear indication of the intra/inter-valley transitions. For example, the green line in Fig. 5c shows the $M_{mn}^{v}$ for LO phonon scattering around the CBM. Clearly, intra-valley scattering is suppressed (around Γ-point), whereas inter-valley scattering is strong (around $g$-point, i.e. the Si $g$-process of scattering in the next Brillouin zone valley). For this transition, only the phonon modes around the second red dot in Fig. 5b participate. This information allows for deeper understanding of transport. In a similar manner, to capture this for the rest of the scattering processes, the form factors, $F_{k,k'}$, can also be computed, similarly to the deformation potentials, only for a few points along high symmetry lines and between valleys. The entire analysis for Si, as well as the result for the mobility calculations which match experiment very well, are all described in detail in Ref. [25], and here we have only provided a summary of this study.

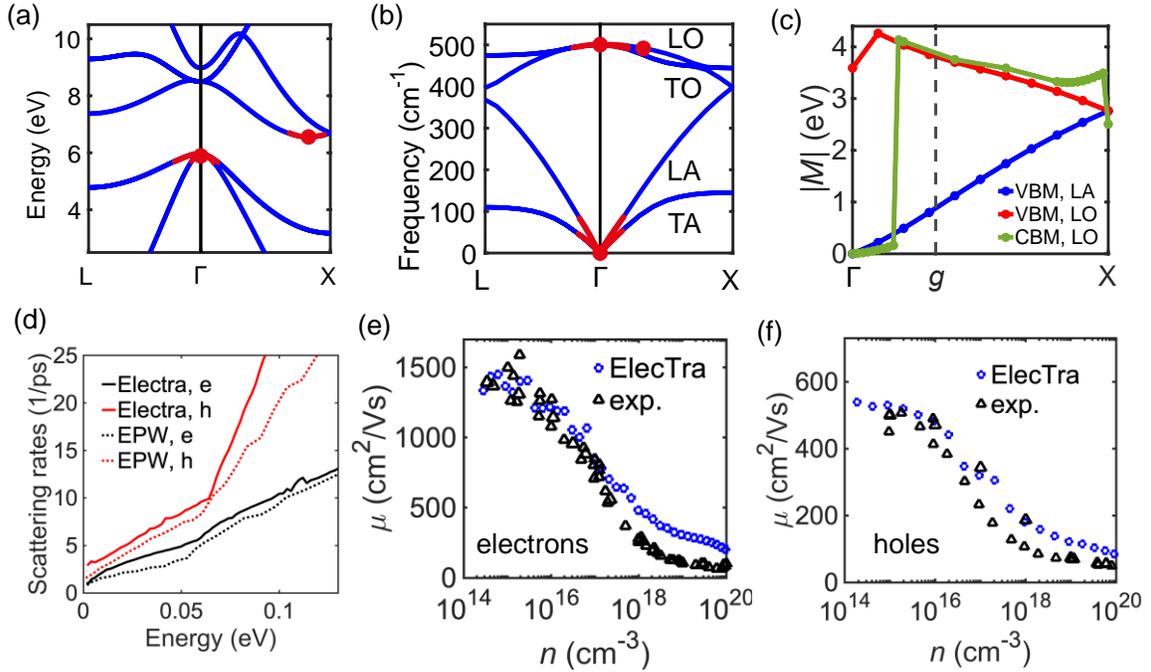

**Figure 5:** (a-b) Si electronic and phononic structures. The red regions indicate the relevant regions from which we extract matrix elements to compute deformation potentials. (c) Some relevant coupling matrix elements for the CB and VB. (d) The scattering rates for electrons (e) and holes (h) from our method (solid lines) and prior EPW results (dashed lines) from Ref. [63]. (e-f) The Si electron and hole mobilities computed with our deformation potential method and the BTE code *ElecTra*, compared to experimental data. Data are from Ref. [64, 65, 66, 67] for electrons, and Ref. [38, 66, 67] for holes. Blue dots are the simulation results and black triangles the experimental data.

**Simulation steps in our method**

The overall simulation steps for the electron-phonon interactions are as follows:

1) Structure relaxation to find the lattice parameter using general DFT codes, such as pw.x in Quantum Espresso, VASP, ABINIT, CASTEP [69]. Quantum Espresso is the favourite first choice we use.

2) Calculation of the wavefunctions using general DFT methods.

3) Calculation of the bandstructures using general DFT codes.

4) Calculation of the phonon spectrum using general DFPT codes such as ph.x in Quantum Espresso; the use of Phonopy [52] with VASP, ABINIT, CASTEP can provide initial guidance and understanding.

5) Find the potential energy change, $\delta_{vq}V(r)$, in DFPT, such as ph.x in Quantum Espresso, ABINIT.

6) Wannier function generation using codes such as epw.x in Quantum Espresso, or Wannier90 [70]. Compare Wannier and DFT bandstructures to validate the generated Wannier functions. (Steps 5/6 only if later the code EPW is used to compute matrix elements)

Up to here, these are the same steps that fully ab initio methods (such as EPW [24]) perform (from here on, those calculations compute >billion matrix elements, usually 100k **k**-points and 10k **q**-points). We continue with the process of extracting deformation potentials using DFPT, which provides the matrix elements, one **q**-point at a time:

7) Find the coordinates of the valleys in the BZ (for CB/VB) and select the extrema points. These will form the initial/final **k**-points for the matrix elements.

8) Find the vectors from each valley to all other valleys, which coincides with the phonon **q**-vectors that will be used.

9) Assign the coordinates of the phonon **q**-points for scattering: i) near the Γ-point for intra-valley scattering, and ii) the **q**-vectors for inter-valley scattering (as in step 8).

10) Use DFPT/Wannier interpolation to compute the matrix elements, $g_{mn}^\nu(\mathbf{k}, \mathbf{q})$, for each phonon process and each valley/band. Using Wannier functions instead of DFT electronic structures, the (interpolated) matrix element calculations are much faster, especially when the VBM/CBM is not located at a high-symmetry point. (In that case in DFT calculations we need a very dense **k**-mesh to cover the VBM/CBM).

11) Derivation of the deformation potentials, $D$, for intra-valley and inter-valley scattering, for acoustic and optical phonons, from $g_{mn}^\nu$.

12) Incorporation of the *D's* within a specially designed database to be read by BTE (previous section) for transport calculations.

The fact that we compute a small amount of matrix elements, makes this method much more computationally affordable compared to fully ab initio Wannier methods, still with accuracy close to those methods (see Ref. [25] for Si). It is on the other hand, it is more expensive compared to other, simpler, deformation potential methods such as the one in Ref. [21], which is designed for reduced computational cost. Thus, our method is a step closer to full ab initio accuracy, with some associated increase in computational cost. Currently it requires some hands-on time to identify the k-vectors and q-vectors for which the deformation potentials are computed, but we are in the process of automating this step.

In Table 1 below we show a computational cost comparison for Si between our deformation potential method (indicated as DP) and DFT+Wannier using the EPW code [24]. The full calculation is presented in our previous work [25]. The calculation starts with the DFPT step, which is common to both methods, typically on a 6×6×6 q-mesh (1$^{st}$ row). Our DP method then computes some matrix elements and then performs Boltzmann transport (2$^{nd}$ and 3$^{rd}$ rows). The corresponding EPW calculation that we have performed (4$^{th}$ row) require significantly more CPU hrs. Note that the 64×64×64 interpolated mesh used is typical for EPW calculations [56], and it can even be an underestimation as the mobility might not be converged for that mesh size. Thus, we save consistently on the generation of the matrix elements and their subsequent use within BTE, which is a cost comparable typically even larger than the coarse-mesh DFPT calculation cost itself. The transport part requires less than 10% and the full calculation less than 20% CPU hours, compared to the fully ab initio method.

Table 1: Comparison between proposed DP method and full ab initio DFT+Wannier with the EPW code [24].

| Step | Parameter | Time (CPU·hours) |
| --- | --- | --- |
| DFPT (common step) | q-mesh = $6 \times 6 \times 6$ | 700 |
| DP extraction (def.pot.) |  | 43 (electrons and holes) |
| *ElecTra* (transport) |  | 384 (using Delaunay triangulation), or 17 (electrons) and 40 (holes) using NN-scan |
| EPW (el-ph + transport) | k-mesh = $64 \times 64 \times 64$<br>q-mesh = $32 \times 32 \times 32$ | 5120 |

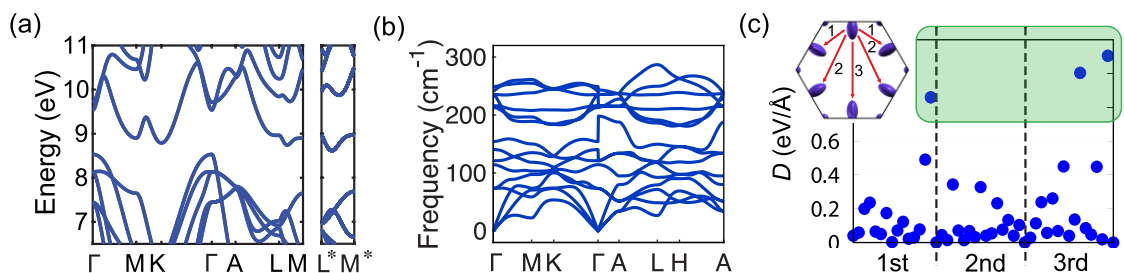

**Figure 6:** (a-b) $Mg_3Sb_2$ electronic and phononic structures. (c) Three types of inter-valley transitions can be identified for each of the six CBM energy pockets around $M^*$ (inset). The deformation potentials caused by each phonon mode for each transition type are shown. Dominant values are in the green box. Inset: The three type of transistions.

It is worth noting that the unit cell of many promising TEs is large and their crystal symmetry low, for full ab initio calculations with practical computation expense. For example, $Mg_3Sb_2$ has five atoms in its unit cell and crystallizes in the trigonal space group, with lower symmetry than cubic. To accurately compute mobility using DFPT is extremely difficult for 3D materials (for 2D is also challenging but feasible [23]). General DFT calculations for $Mg_3Sb_2$ will need ~20 times more CPU hours than for Si. The bandstructure is more complex and many more phonon modes participate (see Fig. 6a-b). Si would need 5k CPU hrs for standard 64×64×64 **k-** and **q**-meshes as shown in Table 1 above. First principle e-ph Wannier calculations would need *~100k* CPU hrs for $Mg_3Sb_2$ (much more time for lower symmetry and larger unit cell). Rather than computing matrix elements throughout the Brillouin zone, the extraction of deformation potentials only needs a limited number of computations (for $Mg_3Sb_2$~50), and will require only *a few thousand* CPU hrs. An example of deformation potential extraction is shown in Fig. 6c. The CBM consists of six equivalent valleys, which results to three different transition types from a certain valley to the other 5 (inset of Fig. 6c). This inter-valley transitions are caused by all phonon modes. In Fig. 6c we compute the deformation potentials caused by each phonon mode individually and for each of the three transition types. Their strength can be quantified, and the dominant ones in each type identified (green box in Fig. 6c). All, or only the dominant ones, can make it to the BTE code for the computation of the transport properties.

Finally, for the rest of the scattering mechanisms, parameters such as $\Delta E_G$ can be computed by calculating the difference between the energy gap of the two constituent materials that form the alloy; $k_\infty$ and $k_s$ are standard outputs of general DFPT codes; and parameters like $\rho$ and $v_s$ can be calculated or found in many databases. $F_{k,k'}$ can also be computed for a few cases within and from valley to valley.

## IV. Transport in nanostructured materials using Monte Carlo

The Monte Carlo (MC) computational formalism we use is described in detail in Ref. [27]. It focuses on the effect of nanostructuring, and for this reason it uses many 'shortcut' deviations from usual MC algorithms to facilitate the challenges that simulating nanostructured geometries present. Specifically, we focus on geometries with the so-called 'hierarchical' nanostructuring, where multiple defects of different nature are introduced in the lattice. Figure 7a-c show typical examples of the geometries: a nanocrystalline geometry with grain boundaries, a nanoporous geometry, and geometry with both grain boundaries and nanopores. Hierarchical nanostructuring is a typical TE material geometry, with the purpose to scatter phonons of different mean-free-paths. However, how these geometries affect electronic transport is much less considered. It gradually becomes important to do so, since large power factors can be achieved when these structured are designed in a specific way [71].

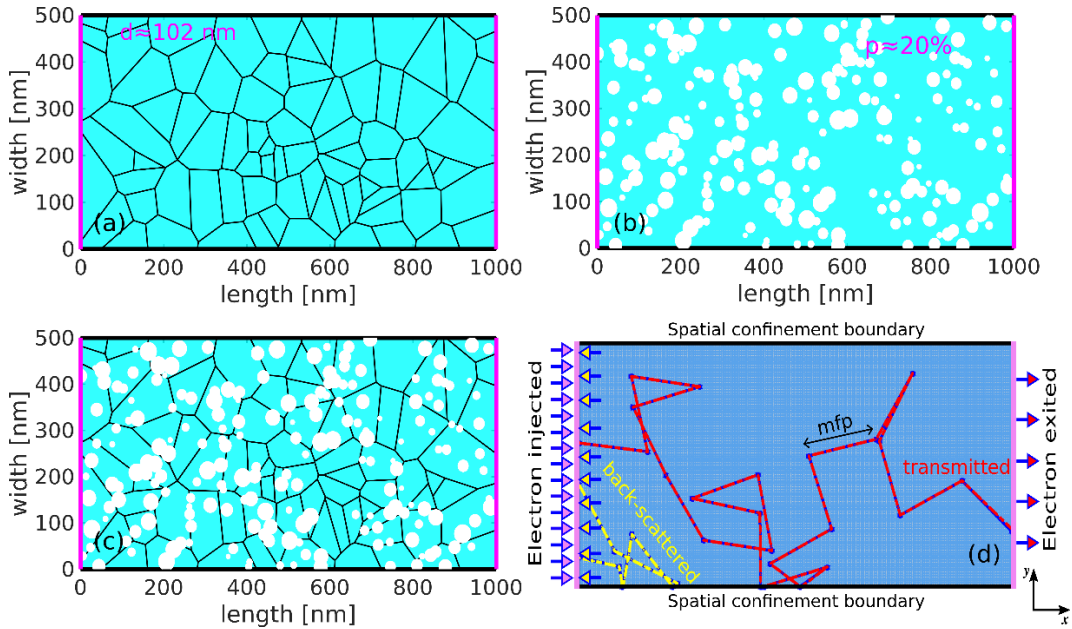

**Figure 7:** A schematic of nanostructured 2D domains populated with (a) grains, (b) pores, and (c) combination of both features to create a hierarchical nanostructure. (d) Ray-tracing of electrons in a pristine domain. An electron moves from the left all the way to the right (red) after traversing several mean free paths (mfp) and consecutive scattering events. The ray-tracing, depicted in yellow, shows backscattering to the left side of the domain that does not contribute to the flux calculation.

For computational simplicity and speed reasons, we follow two directions in using the real material properties and ab initio information extracted in the previous sections: 1) We consider parabolic bands and map the complexity of the real material bandstructure onto parabolic bands. For this we extract a conductivity and DOS effective masses ($m_C$ and $m_{DOS}$, respectively) for the real bandstructure. We also use as input the deformation potentials to form the scattering rate equations. 2) In the second method, we extract the DOS(E), band velocity, $v$(E), mean-free-path $\lambda$(E), and scattering times, $\tau$(E), for the material of interest using the *ElecTra* BTE code (which already uses the deformation potentials).

We begin below by describing the approach we follow to extract the effective masses to be used in MC, which is presented in [29, 72], and then we describe the MC algorithm we employ, with specific modifications and peculiarities compared to current algorithms, designed specifically for the numerical complexities introduced by nanostructuring.

**Effective mass extraction:**

We compute the $m_{DOS}$ as the effective mass of an isotropic parabolic band that gives the same carrier density as the actual band structure. We evaluate $m_C$ as the effective mass of an isotropic parabolic band which maps the average velocity of the band states weighted by their contribution to transport. For this, we employ a simple ballistic field effect transistor model, extract the average injection velocity in the sub-threshold regime, and map that velocity to a parabolic band which provides the same injection velocity. The process is as follows (using the conduction band as an example): We consider the non-degenerate regime, in which the carrier density $n$ can be expressed as:

$$n = N_C \cdot \exp\left(\frac{E_0 - E_f}{k_B T}\right) \tag{7}$$

where $E_0$ is the energy of the band edge, $E_F$ is the Fermi level, $k_B$ is the Boltzmann constant, $T$ is the temperature, and $N_C$ is the effective density of states calculated as:

$$N_C = 2\left(\frac{m_{DOS} k_B T}{2\pi \hbar^2}\right)^{3/2} \tag{8}$$

For a generic numerical band structure, the carrier density $n$ can be calculated as:

$$n = \frac{2}{(2\pi)^3}\Sigma_{k,n} f_{E_{(k,n)}} dV_k \qquad (9)$$

where the sum is over all *k*-points and bands in the first Brillouin zone, $f_{E_{(k,n)}}$ is the Fermi-Dirac distribution, and $dV_k$ is the volume element in $k$ space, which usually depends only on the mesh. Then the $m_{DOS}$ can be obtained by combining Equations (7-9).

The conductivity effective mass in a specific direction, $m_{C,i}$, is calculated from the injection velocity $v_{inj}$ of the carriers in the sub-threshold regime of a field effect transistor (FET):

$$m_{C,i} = \frac{2k_B T}{\pi v_{inj}^2} \qquad (10)$$

where the injection velocity is extracted as the FET current divided by the injected charge at the source contact:

$$v_{inj} = \frac{I_{FET}}{\frac{q}{2\pi^3}\Sigma_{k,n} f(E_{k,n}-E_{F,S}) dV_k} \qquad (11)$$

where $E_{F,S}$ is the source Fermi level, and $I_{FET}$ is the FET current density, which is computed from the difference between the source and drain currents. We assume that the drain current is negligible under high drain voltage, thus $I_{FET}$ can be given by:

$$I_{FET} = \frac{e}{2\pi^3}\Sigma_{\boldsymbol{k},n} f_{(E_{k,n}-E_{F,S})} |v_{k,n}| dV_k \qquad (12)$$

The $v_{inj}$ is identified and the conductivity effective mass is extracted from Equation (10) along the three Cartesian directions. Thus, three conductivity effective masses are extracted, and the final conductivity effective mass is calculated by averaging as:

$$m_C = \frac{3}{\frac{1}{m_{C,x}}+\frac{1}{m_{C,y}}+\frac{1}{m_{C,z}}} \qquad (13)$$

**<u>Monte Carlo method for nanostructures</u>**

Monte Carlo simulations involve the ray-tracing of particle trajectories rather than the direct solution of partial differential equations. These particles are allowed to move in the domain in both left/right directions under the influence of a driving force, and the net flux is computed in a statistical manner. Although this method served well simulations in bulk materials with success over many years, for nanostructured materials large difficulties are encountered, which make simulations computationally extremely expensive and logistically cumbersome. The main difficulty that arises for MC methods for nanostructures, is that the presence of nanostructuring features such as grain boundaries, pores, nanoinclusions, potential barriers, etc. to name a few, reduce the flux in the domain at such a degree, which makes it very difficult to gather enough statistics for converged flux results. This is particularly difficult under linear response (either under a small voltage or temperature difference), where the two opposite going fluxes vary only slightly. Typical simulations in the literature require from 10s of thousands to even millions of trajectories for adequate results [73, 74, 75, 76].

Another numerical peculiarity that is encountered in MC simulations for low energy electrons even in pristine materials, makes the computational difficulty even larger. Noticeably, the low energy carriers near the band edge have small velocities. Thus, there exists a population of slow moving electrons which ends up dominating the computation time (even so in certain cases where the Fermi level is placed higher into the bands). To make things worse, in the presence of potential barriers, the contribution of those low-energy electrons to transport is insignificant, but requires unnecessarily large computational resources.

In order to address the difficulties described above and enable efficient and accurate MC simulations tailored to nanostructures, we have developed a hybrid MC algorithm which: i) merges information from analytical BTE solutions with the numerically extracted flux, ii) considers a single flux initialized from the left only and injected into the channel where it is ray-traced to either of the contacts, iii) does not require the application of a driving force. The method we present provides the same accuracy as common methods, but with a significantly reduced computational cost. It has many differences and peculiarities compared to standard MC methods, to tackle the issue of computational complexity. Details are provided in [27], and in this paper we provide the essentials and focus on the coupling and use of the various ab initio parameters within MC.

We use the incident flux (single-particle) approach, where the electrons are initialized at the domain boundaries one-by-one and propagate until they exit at either boundary (propagated to the other side or back-scattered as shown in Fig. 7d). Here we consider a two dimensional domain for numerical simplicity. Regarding the large computation time associated with the ray-tracing of low energy electrons, we consider a mean-free-path (mfp) approach, rather than the picking of random free-flight time and the self-scattering approach as is common practice [30, 77]. We compute the total scattering rate of the particle using the deformation potential expressions described earlier, and using its bandstructure velocity we calculate its mean-free-path, $\lambda(E)$. The particle propagates one mfp at a time (as in Fig. 7d), and then undergoes (enforced) scattering. In the case of acoustic phonon scattering for 3D parabolic bands, for example, the mean-free-path is constant in energy. Thus, electrons at all energies are treated in the same way, with only different free-flight durations. However, a tabulated $\lambda(E)$ can also be employed directly as computed above by *ElecTra*.

**The simulation procedure is as follows:** Following the incident-flux method, we initialize and inject electrons in the channel domain, but only from the left side, and neglect the injection of flux from the right of the channel. Thus, we refer to this as the `single-flux' method. We initialize those electrons uniformly in energy, rather than according to their density of states (DOS), as in typical MC methods. Then one-by-one the electrons are ray-traced, by alternating between free-flight events of a mfp distance and intermediate scattering events. The upper/lower closed boundaries simply specularly reflect back the electrons in the domain.

To gather the flux statistics, we record the time spent in the domain by those electrons which propagate all the way from the left to the right end of the domain and exit from there, referred to as its time-of-flight (ToF). All electrons that are back-scattered to the left do not contribute to the flux and are not considered. The average ToF per particle is then computed as:

$$<ToF(E)> \;=\; \frac{\sum t_r(E)}{N_r(E)} \tag{14}$$

where, $t_r(E)$ is the ToF for a single electron, and $N_r(E)$ is the number of electrons that make it to the right end. We chose to keep the energy dependence since we only deal with elastic transport conditions for the example we present here. Then the averaged <ToF(E)> is used to calculate the flux per simulated electron at each energy as:

$$F(E) = \frac{1}{<ToF(E)>} \tag{15}$$

Using the flux per electron at a given energy, we can form the overall flux in energy by multiplying by the density of states (DOS), $g(E)$, either in its parabolic form, or its tabulated form from *ElecTra* for the real complex band material, which essentially is proportional to the transport distribution function (TDF) of the analytical BTE as:

$$\Xi(E) = C\, F(E)\, g(E) \tag{16}$$

This is simply because the product of flux with DOS provides essentially the flow of charge, which is directly related to conductivity in the same way the TDF determines the conductivity. The proportionality constant $C$ in the equation of the TDF accounts for the super-electron charge that is typically used in MC (the fact that we only simulate a finite number of electrons), and geometrical factors related to the simulation in a finite 2D domain rather than an infinite 3D domain (and connects the conductance to conductivity). We use $C$ to map the MC simulated TDF to the TDF that can be derived and extracted from the analytical solution of the BTE (or the one extracted from *ElecTra*), for the case of pristine material alone. To extract $C$, we calculate the electrical conductivity as a function of the Fermi energy both in the linearized BTE and with MC by integrating the TDF times the Fermi derivative in energy. We then find the mapping factor $C_{EF}$ for the conductivity at each Fermi energy. This comes to be almost constant for all $E_F$, so we take the average of those values as the overall final $C$. After this, it also turns out that the TDFs from analytical BTE and MC are almost identical (see Fig. 8a).

After obtaining the transport distribution function numerically from MC and for the MC TDF, $\Xi_{MC}(E)$ by calibrating to the linearized BTE one, we can substitute $\Xi_{MC}(E)$ in the place of the analytical $\Xi(E)$, which is given by $\tau(E)v^2(E)g(E)$ in the usual BTE formulas below. The electron conductivity is then calculated as:

$$\sigma = q_0^2 \int_E \Xi_{MC}(E)\left(-\frac{\partial f}{\partial E}\right) dE \tag{17}$$

and the Seebeck coefficient as:

$$S = \frac{qk_B}{\sigma} \int_E \Xi_{MC}(E) \left(-\frac{\partial f}{\partial E}\right) \left(\frac{E - E_f}{k_B T}\right) dE \tag{18}$$

Further, the TE power factor and the electronic thermal conductivity are evaluated, respectively, as:

$$PF = \sigma S^2 \tag{19}$$

$$k_{el} = \frac{1}{T} \int_E \Xi_{MC}(E) \left(-\frac{\partial f}{\partial E}\right) (E - E_f)^2 dE - \sigma S^2 T \tag{20}$$

The above transport coefficients are computed using MC initially for the pristine material configuration, where the calibration of the constant *C* takes place. The electronic conductivities from the analytical BTE and the simulated MC in this way, are again very similar, almost identical (see **Fig. 8c**). The idea is that once this is calibrated, we can perform MC simulations for complex nanostructured domains without further calibration and benefit from the robustness of the single-flux injection method.

Thus, by using the MC extracted $\Xi_{MC}(E)$, and by multiplying it by the derivative of the Fermi distribution function with respect to energy (i.e. $\partial_E f = \frac{\partial f}{\partial E}$) for the electrical conductivity (see Fig. 8b), and with respect to temperature (i.e. $\partial_T f = \frac{E - E_f}{k_B T} \times \frac{\partial f}{\partial E}$) for the Seebeck coefficient calculations, we essentially account for linear response. In this way we effectively eliminate: i) the need for two counter propagating simulation fluxes (now this is captured by the derivatives), and ii) the need for an application of a driving force, either a voltage difference or a temperature difference. Thus, we avoid the peculiar situation where a small enough potential difference window does not provide enough statistics, while a large enough could make the simulation deviate from linear response or from the range of voltages that TE materials utilize. Note also that the acquisition of adequate statistics is even more difficult in the case of a temperature gradient for the Seebeck coefficient in common bi-directional flux methods. We do not only need to differentiate between the right and left going fluxes, but also the ones which flow above and below the Fermi level.

We now consider the effect of nanostructuring. We simulate the geometries shown in Fig. 7 above and compute the electrical conductivity and power factor, shown in Fig. 9a and Fig. 9b, respectively. We show the conductivity as a function of the Fermi level, $E_F$, for the cases of the pristine channel (blue line), the channel with grain boundaries of approximately 100 nm size (yellow line), the porous geometry with p=20% porosity (purple line), and the geometry with the combination of grain boundaries and pores (red line). By the red-dashed line we show the conductivity computed using Matthiessens's rule (by combining the result for the grain boundaries and the pores analytically), which matches the MC line well, indicating that our simulation method provides the expected results.

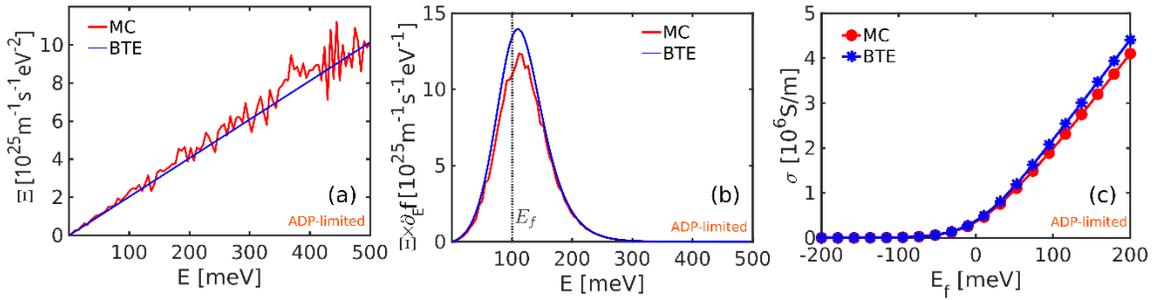

**Figure 8:** (a) Transport quantities for a pristine domain calculated using the analytical BTE (blue) and MC formalism (red). (a) The transport distribution functions versus energy. (b) The product of the transport distribution function product with $\partial f/\partial E$ versus energy. (c) Electrical conductivity vs the Fermi energy. The calculations consider only acoustic phonon scattering.

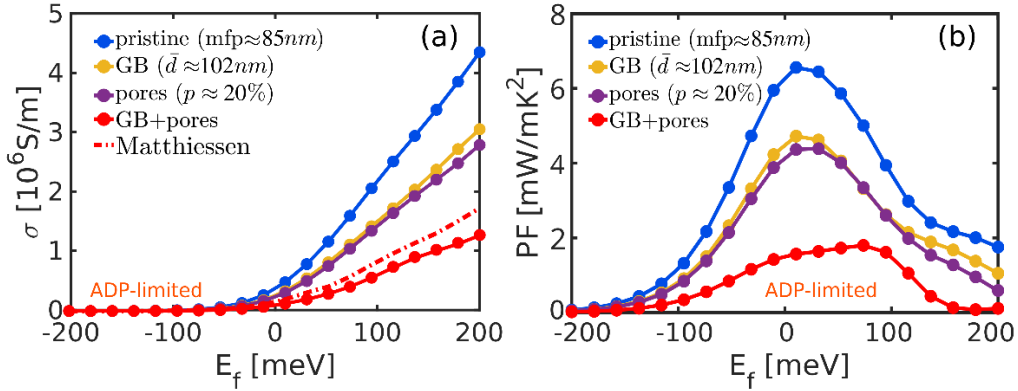

**Figure 9:** (a) Electrical conductivity and (b) power factor (PF) computed from the MC formalism in nanostructured domains populated with grain boundaries (yellow), with pores (purple), and the combination of them (red) as a function of Fermi energy ($E_f$). The dotted line in (a) shows the calculated electrical conductivity using Matthiessen's rule. Only acoustic phonon scattering is considered in the calculations. The blue lines indicate the pristine material properties.

# V. Conclusions

In this paper, we provided a review of a computational methodology which allows the extraction of the electronic and thermoelectric coefficients of complex nanostructured materials. The method merges three distinctive parts: i) a Boltzmann Transport Equation solver that we have recently developed, named *ElecTra*, which takes as input a complex bandstructure and scattering parameters, and provides transport coefficients and relevant quantities; ii) an ab initio methodology to extract the scattering parameters needed by *ElecTra*, and in particular deformation potentials; and finally iii) a Monte Carlo simulator which can utilize transport quantities from the ab initio calculations and the BTE transport above, with many features that deviate from common methods, specifically designed to provide robust computation for nanostructured simulation domains. In fact, each of the three methods we describe utilize unconventional routes that allow for faster and more robust calculations, without compromising accuracy at a significant degree. We believe this multi-physics framework, but also each of the simulators individually, are truly enabling, and can contribute significantly in exploring the electronic and thermoelectric properties of pristine and nanostructured materials in an efficient manner.

# Statements and Declarations

## Funding

This work has received funding from the European Research Council (ERC) under the European Union's Horizon 2020 research and innovation programme (Grant Agreement No. 678763), the European Union's Horizon 2020 research and innovation program under grant agreement No. 863222 (UncorrelaTEd), and the European Commission under the Grant agreement 788465 (GENESIS).

## Competing interests

The authors have no relevant financial or non-financial interests to disclose.

## Author Contributions

All authors contributed to the study conception and design. Material preparation, data collection and analysis for section II (BTE) was performed by Patrizio Graziosi. Material preparation, data collection and analysis for section III (ab initio calculations) was performed by Zhen Li. Material preparation, data collection and analysis for section IV (Monte Carlo) was performed by Pankaj Priyadarshi. The overall supervision and draft of the paper was performed by Neophytos Neophytou. All authors read and approved the final manuscript.

## Data availability

Data can be shared after reasonable request to the authors. The software used to produce the results is publicly available.

*ElecTra*: https://zenodo.org/record/5074944#.Y-DT6HbP2Uk.

*Monte Carlo*: https://warwick.ac.uk/fac/sci/eng/research/grouplist/sensorsanddevices/computational_nanotechnology_lab/uncorrelated/